\documentclass[%
 reprint,
superscriptaddress,
twocolumn,
amsmath,
amssymb,
aps,
prl,
longbibliography,
]{revtex4-2}
\usepackage{graphicx}
\usepackage{dcolumn}
\usepackage{bm}
\usepackage{hyperref}
\usepackage{amsmath}  
\usepackage{amssymb}
\usepackage{xcolor}
\usepackage{braket}
\begin{document}

\title{Density functional theory based investigation of heavy fermion band candidates in triplet superconductor UTe$_2$}

\date{June 2024}
\author{Shouzheng Liu}
\email{sl7956@nyu.edu}
\author{L. Andrew Wray}
\email{landrewwray@gmail.com}
\affiliation{Department of Physics, New York University, New York, New York 10003, USA}

\begin{abstract}
The compound UTe$_2$ is of great recent interest as a spin triplet superconductor that is thought to realize a topologically nontrivial superconducting order parameter. Though UTe$_2$ is considered to be a heavy fermion material, only light-electron Fermi surfaces have been compellingly identified, while a possible heavy electron Fermi surface near the momentum space Z-point has been the subject of more speculative and incongruent interpretations based on density functional theory (DFT+U) and experimental measurements. Here we explore the DFT+U band structure within a tight binding model framework to identify emergence mechanisms for heavy fermion Fermi surfaces. Applying a Gutzweiller-like renormalization term to $f$-electron kinetics is found to distort the lowest energy $f$ electron band from a saddle point to a local minimum at the Z-point, introducing a shallow Fermi pocket and greatly improving compatibility with recent experiments.
\end{abstract}
\maketitle

Spin-triplet p-wave superconductors are promising platforms for topological quantum computation through surface state Majorana modes \cite{majorana_pwaveSC1,majorana_pwaveSC2,majorana}. Overwhelming evidence for triplet pairing has been found in UTe$_2$ ($T_c\sim$2K), including an upper critical field that exceeds the Pauli limit for all three crystallographic axes \cite{RanSheng_EarlyPaper,RanSheng_LargeHc2} and Knight shift measurements inconsistent with singlet Cooper pairing \cite{RanSheng_EarlyPaper, NMR_tripletPairing}. However, the strongly correlated electronic structure that underlies local pairing interactions remains poorly understood. Heavy 5$f$ bands have poor spectral visibility for electron probing techniques such as angle resolved photoemission (ARPES) \cite{Gutzwiller, Gutzwiller_appro} and interpretations of the 5$f$ band structure of UTe$_2$ are inconsistent and speculative. Much attention has focused on the momentum space Z-point, where ARPES has tentatively attributed a heavy $f$-electron Fermi surface \cite{LinMiao_ARPES}, and where the primary 5$f$ Fermi surface candidate is found in density functional theory (DFT+U) simulations for which Hubbard U is tuned to reproduce the quasi-2D light-electron band structure of UTe$_2$ \cite{Yanase_DFT1,Yang_DFTandDMFT}. Here, we examine the candidate 5$f$-electron band structure within a tight binding model obtained from DFT+U calculations. When hopping coefficients are renormalized to represent the effect of strong correlations, an electron-like Fermi surface emerges at the Z-point due to hybridization from a significantly dispersive higher energy band. This in turn enables qualitatively improved correspondence with transport characterization and quantum oscillation (QO) measurements of the Fermi surface contours.

\begin{figure}[]
    \centering
    \includegraphics[width=8.7cm, trim={0.5cm 0 0 0},clip]{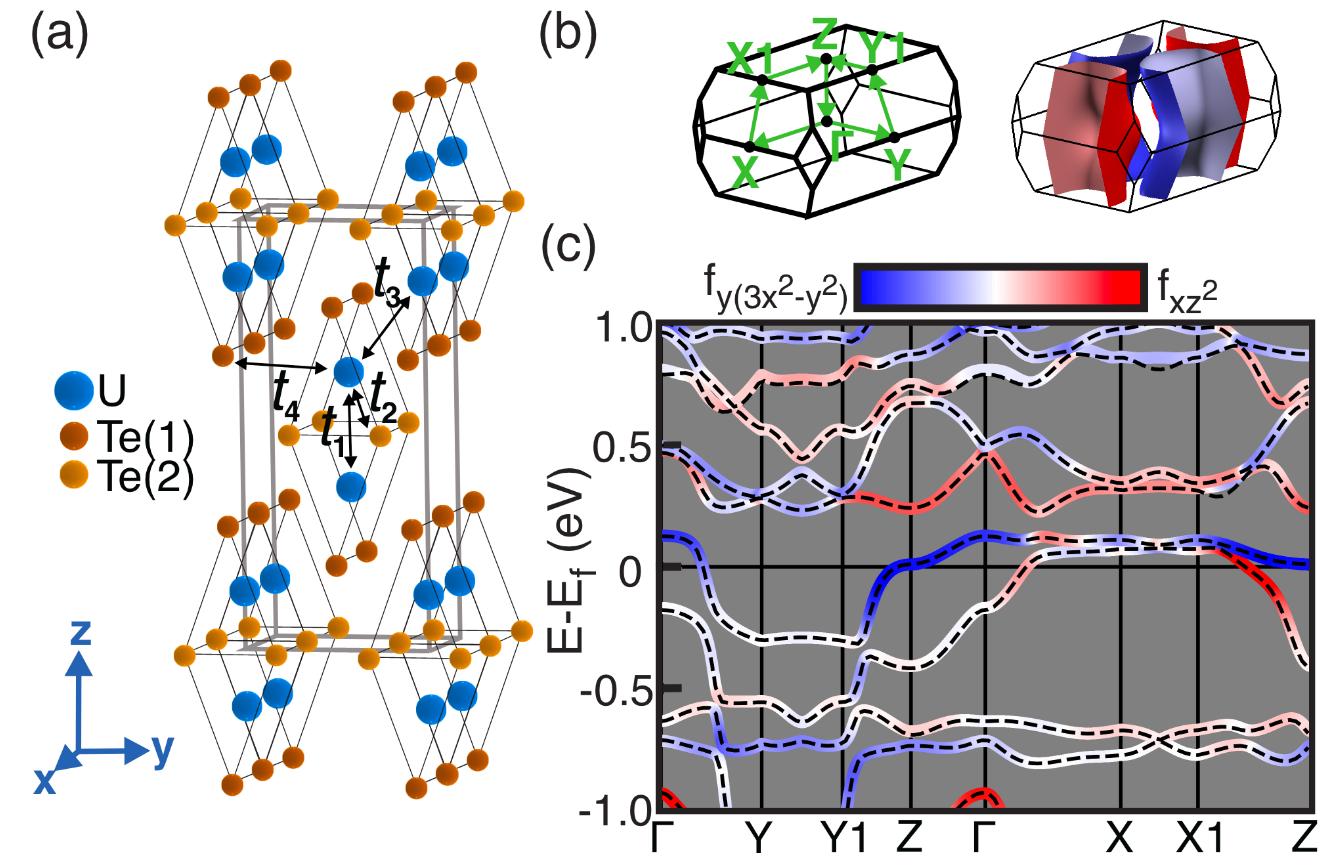}
    \caption{\textbf{DFT+U band structure.} (a) Lattice structure of UTe$_2$, with typical uranium $f$-electron hoppings indicated by black arrows. (b) First Brillouin zone of UTe$_2$ and Fermi surfaces from the tight-binding (TB) model. (c) Band structure of UTe$_2$ obtained from DFT+U (dashed line) and from the TB model (colored line). Shading indicates relative orbital weight between $f_{xz^2}$ (red) and $f_{y(3x^2-y^2)}$ (blue). }
    \label{fig:fig1}
\end{figure}

The light-electron band structure of UTe$_2$ includes two quasi-1D Fermi sheets (Fig. \ref{fig:fig1}(b-c)) that disperse along the $k_x$ and $k_y$ momentum axes (see axes in Fig. \ref{fig:fig1}(a)) and are associated with the 6$d$ and Te-5$p$ orbitals, respectively. These bands are well described by first principles-based modeling and correspond with prominent features in ARPES and QO investigations \cite{LinMiao_ARPES, Yanase_DFT1,Yang_DFTandDMFT,aoki_QO}. Strong but poorly resolved photoemission intensity has also been observed surrounding the Z-point \cite{LinMiao_ARPES} with $k_z$ axis momentum dependence that may be critical to explaining the highly 3D conductivity of UTe$_2$ \cite{YunSukEo_cAxisTransport} if the feature is designated as a band. High resolution QO measurements along the $\hat{c}$ axis ($\hat{z}$) show a low-frequency peak (\textless0.5 kT), and angle-dependent QO data suggest that this signal could be produced by a small, largely isotropic Fermi pocket with dimensions similar to the ARPES Z-point feature \cite{aoki_QO2,RanSheng_QO}. Another recent QO survey lacks such a peak \cite{Eaton_QO}, however this may be due to the use of a torque-based method, which is insensitive to isotropic Fermi surfaces. Furthermore, the temperature dependence of thermoelectric power (S) indicates a Lifshitz transition under an applied field, consistent with a shallow ($\sim$2 meV) heavy-electron Fermi pocket \cite{NiuQian_FSInstability}. 

These features have proven challenging to reproduce in first principles-based models. Fine-tuned DFT+U calculations can place a 5$f$ band at the Z-point, but it forms a saddle point rather than a true electron pocket as implied by QO \cite{Yanase_TB,RanSheng_QO}. Additionally, DMFT-based investigations have proposed the formation of a small Fermi pocket around the $\Gamma$ point \cite{DMFT_LowTpocketFS} or Z-point \cite{DMFT_Zpocket} at low temperature, however these calculations significantly distort the two sets of quasi-1D Fermi surfaces and are thus questionable as a basis for explaining the high-frequency QO branches. 
\par
The UTe$_2$ lattice is orthorhombic and belongs to the Immm space group \cite{UTe2_stru}. Each unit cell contains two uranium atoms that form a dimer, encapsulated by a diamond-shaped cage of tellurium atoms (Fig. \ref{fig:fig1}(a)). The U-dimers are arranged in a ladder-like structure along the $\hat{x}$ axis and the Te(2) atoms form a tightly-spaced chain along the $\hat{y}$ axis, resulting in the two quasi-1D light-band Fermi sheets that have been widely identified in ARPES, QO, and DMRG investigations, and are the only Fermi surface features present in the Fig. \ref{fig:fig1}(c) DFT+U band structure. We performed DFT+U calculations on UTe$_2$ using a $16\times16\times16$ k-mesh with WIEN2k, applying a moderate Hubbard U = 1.6 eV that is consistent with earlier studies \cite{Yanase_DFT1,wien2k}. A tight-binding (TB) model that reproduces the low energy DFT+U band structure (compare dashes and shaded lines in Fig. \ref{fig:fig1}(c)) is obtained by down-folding the WIEN2k calculations into U-5$f$, U-6$d$, Te-5$s$ and Te-5$p$ orbitals and applying a cutoff to all hoppings longer than 16 \AA. The SKEAF software package \cite{SKEAF_QO_simu} is used to obtain QO curves from simulated band structure, and semi-classical transport calculations were performed using the LinReTraCe package \cite{Linretrace}.
\par
First principles DFT calculations approximate the correlation energy using exchange-correlation functionals that depend on electron density. This approach is effective in weakly correlated systems where electrons exhibit free-electron behavior, but the Kohn-Sham basis of DFT does not accurately describe strongly correlated electronic states. In the DFT+U method, a local mean field interaction term is applied to strongly correlated orbitals, with the effect of adding an orbital-resolved energy shift ($\Delta\epsilon_i$) that can be represented as an on-diagonal term for a tight binding model with an appropriately chosen orbital basis \cite{Gutzwiller_appro}:
\begin{equation}
    H_{DFT+U}=\sum_{i\neq j}t_{i,j}c_{i}^{\dagger}c_{j} + \sum_{i}(\epsilon_i+\Delta\epsilon_i)c_{i}^{\dagger}c_{i}.
\end{equation}
This can align simulations with band features such as charge transfer gaps but remains intrinsically inaccurate for band dispersion and density of states (DOS).  The Gutzwiller approximation goes a step further by introducing a variational parameter $z_i$ that represents the fraction of an electron available after multiplet splitting or any other loss of local degrees of freedom due to strong correlations \cite{Gutzwiller}. In the infinite dimensional limit, this representation leads to an effective one-particle Hamiltonian that modifies not just band energies, but also electron kinetics \cite{Gutzwiller_appro}:
\begin{equation}
    H_{G}=\sum_{i\neq j}t_{i,j}z_iz_jc_{i}^{\dagger}c_{j} + 
    \sum_{i}(\epsilon_i+\Delta\epsilon'_i)c_{i}^{\dagger}c_{i}
\end{equation}
The variational term is bounded as $0\leq z_i\leq 1$, and also factors into the conversion from band structure to DOS. As such, it directly reduces both band dispersion and the visibility of the corresponding band features for electron counting spectroscopies such as ARPES and STM.
\par
Motivated by Gutzwiller's approximation, we introduce a dispersion correction parameter $\alpha$ such that U-5$f$ creation and annihilation operators are renormalized by a factor of $\sqrt{\alpha}$ (analogous to $z_i$) in our TB model. This results in a scaling factor of $\alpha$ for all inter-cell $f\leftrightarrow f$ hoppings and $\sqrt{\alpha}$ for all $f\leftrightarrow \text{non-}f$ hoppings. Values of alpha are considered in the range $0.15 \leq \alpha \leq 0.4$, a typical renormalization range for multiplet-split $f$-electrons (\emph{cf.} \cite{Gutzwiller_smb6}). Local Hamiltonian terms that define orbital energies are left unrenormalized for simplicity, as this eliminates the need for an additional \emph{ad hoc} energy term and does not change the realized Fermi surface topologies (see Fig. S2-S3 of the Supplementary Material, SM).

\begin{figure}[ht]
    \centering
    \includegraphics[width=0.48\textwidth]{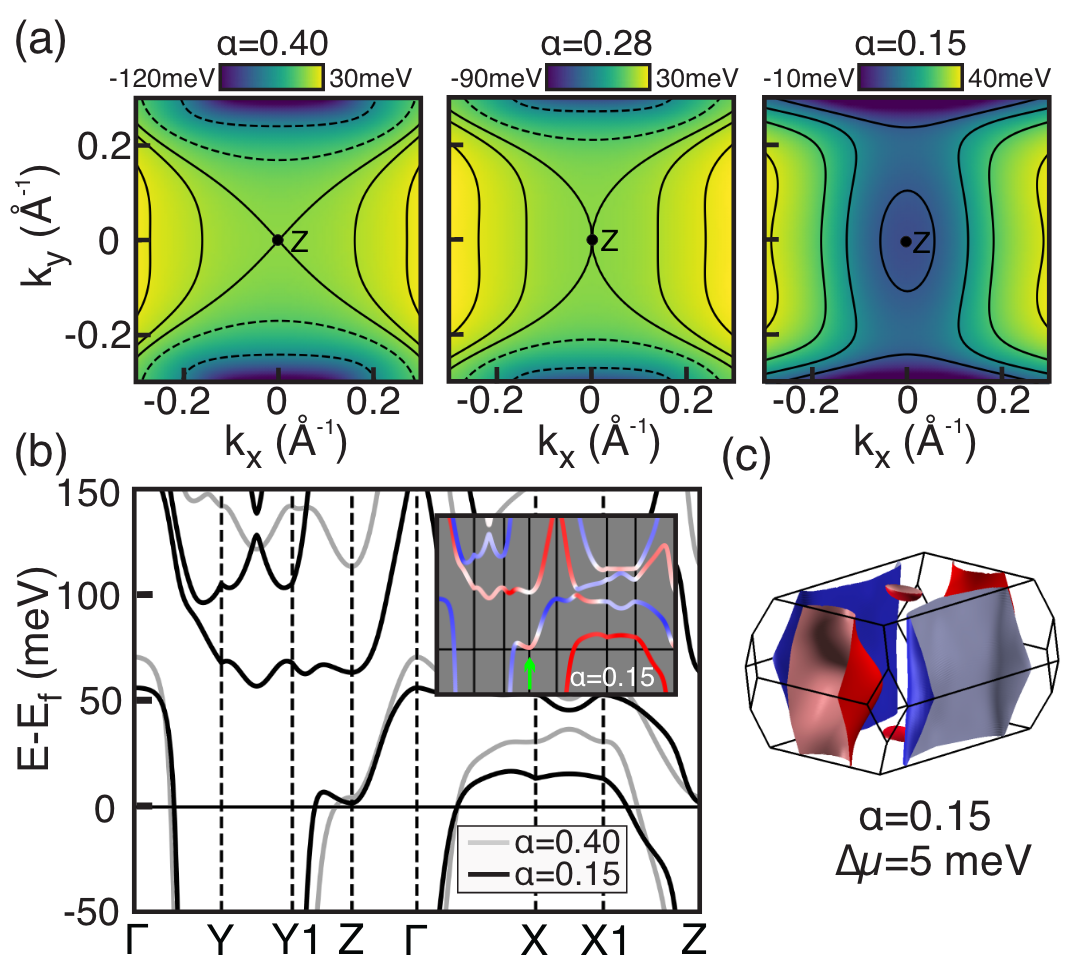} 
    \caption{\textbf{Emergence of a Z-pocket.} (a) Band energy map in the $k_z = \frac{2\pi}{c}$ plane showing constant energy contours for various $\alpha$ values. (b) Evolution of the band structure near the Fermi energy, showing the emergence of a local minimum at the Z point. Inset shows the orbital symmetry projection at $\alpha = 0.15$. (c) Fermi surfaces obtained from the TB model with a renormalization parameter $\alpha = 0.15$. Chemical potential is shifted by 5 meV to align the Fermi surface dimensions with QO.}
    \label{fig:fig2}
\end{figure}
\par

Band symmetries near the Fermi level feature significant components from the 5$f_{y(3x^2-y^2)}$ and 5$f_{xz^2}$ orbitals (see shading in Fig. \ref{fig:fig1}(c)), consistent with angular momentum ($m_j$-basis) symmetry attributions in a recent DMFT investigation \cite{DMFT_OrbitalSelectiveKondo}. A band closely associated with the 5$f_{y(3x^2-y^2)}$ orbital can be seen immediately above the Fermi level at the Z-point, with a saddle point dispersion that curves downward along $k_y$ and upward along the other two principal momentum axes. An electron-like 5$f_{xz^2}$ band with large dispersion along the $k_x$ and $k_z$ axes is found 0.2 eV higher in energy. The combination of large dispersion in the 5$f_{xz^2}$ band and strong symmetry-allowed hybridization between the two bands provides a mechanism for the emergence of a heavy electron Z-pocket. 

Applying the renormalization parameter $\alpha$\textless 1 causes these 5$f$ bands to draw together and come down in energy due to the loss of level repulsion with Te valence bands. Convergence of the bands in turn leads to symmetry exchange, and the large electron-like dispersion of the 5$f_{xz^2}$ band imparts upward parabolicity on the lower 5$f_{y(3x^2-y^2)}$ band (see arrow in Fig.\ref{fig:fig2}(b) inset). Dispersion at the Z-point is converted from a saddle point to a local minimum beneath a critical value of $\alpha \leq 0.28$. A high order saddle point (HOSP) with the form $k_x^2-k_y^4+k_z^2$ is realized at the critical point (see Fig. \ref{fig:fig2}(a)), constituting a high-order van Hove singularity (HOVHS) with multiplicity 3 \cite{HOVHS_classification} that will result in a step-like DOS distribution. Regardless of the precise value of $\alpha$, the proximity to a HOSP tends to enhance DOS by confining dispersion along the $k_y$ axis. Recent studies suggest that some HOSPs can lead to a faster-than-logarithmic divergence in DOS for 3D materials and may strengthen Cooper pairing interactions within cuprates \cite{HOVHS}. A value of $\alpha$=0.15 is adopted to explore the electron-like dispersion regime, with the chemical potential shifted up by 5 meV to achieve Fermi contours that approximately match the Z-point ARPES feature \cite{LinMiao_ARPES} (see Fig. \ref{fig:fig2}(c) Fermi surface map).

\begin{figure}[ht]
    \centering
    \includegraphics[width=0.48\textwidth]{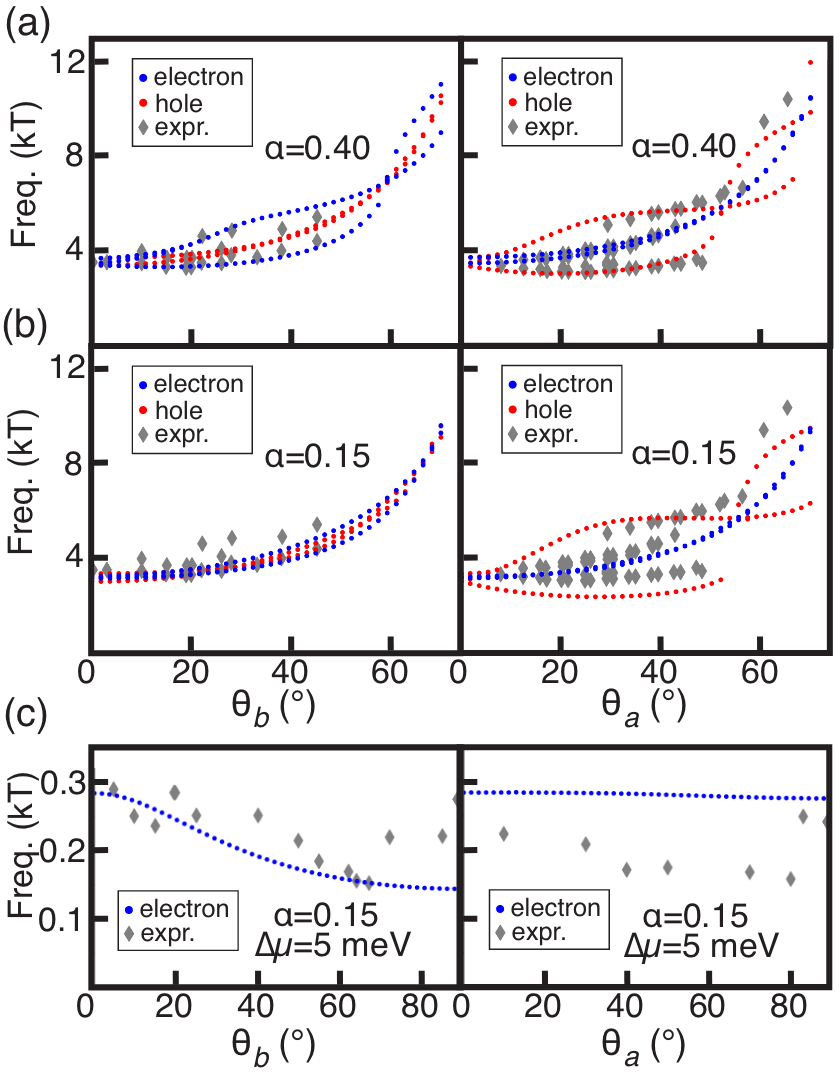} 
    \caption{\textbf{Angular dependence of QO frequencies.} Calculations with renormalization parameter (a) $\alpha = 0.40$ and (b) $\alpha = 0.15$ are compared with experimental data. Experimental data in the $\hat{c}/\hat{b}$ plane are from Ref. \cite{Eaton_QO} and $\hat{c}/\hat{a}$ plane data are from Ref. \cite{aoki_QO}. (c) Angular dependence of QO with renormalization parameter $\alpha = 0.15$ and chemical potential shifted by 5 meV, compared with data from Ref. \cite{RanSheng_QO}.}
    \label{fig:fig3}
\end{figure}
\par

The angular dependence of QO measurements is defined by Fermi surface contours and can be used to assess the accuracy of band structure predictions. Previous DFT+U calculations with U\textgreater 1.6 eV have yielded significant correspondences but incomplete agreement with experimental QO frequencies. These calculations exhibit monotonically increasing QO frequencies when rotating away from crystallographic $\hat{c}$ axis, as the quasi-1D Fermi sheets hybridize to yield tube-like Fermi surfaces parallel to the $\hat{c}$ axis. Level repulsion with heavy U-5$f$ electrons distorts these tubes, causing splitting in the QO curves \cite{aoki_QO}. This is initially strengthened by turning on renormalization, due to the convergence of 5$f$ bands towards the Fermi level as discussed above. Using our `moderate' renormalization value of $\alpha = 0.4$, we find a surprisingly close overlay between the simulation and the cluster of experimental QO branch dispersions at $\sim4$ kT (Fig. \ref{fig:fig3}(a)). The experimental signal-to-noise ratio is relatively low for $\theta_a$\textgreater 60\textdegree, which may influence the apparent correspondence between experiment and simulation results in this region. One should also note that using QO to fine tune parameters is a dubious exercise given the error intrinsic to DFT-based approaches in this regime of strong correlations.

Further strengthening renormalization from $\alpha = 0.4$ to $0.15$ preserves the same basic picture, though it disrupts the close correspondence of experiment and theory for the high frequency $\sim4$ kT QO branches at a quantitative level (Fig. \ref{fig:fig3}(b)). More significantly, the emergence of a local band minimum at the Z-point for $\alpha < 0.28$ creates a small Fermi pocket, fulfilling a key criterion for establishing a 1:1 feature correspondence with recent QO measurements that show a low frequency $\sim 0.2$ kT QO branch \cite{RanSheng_QO} (see Fig.\ref{fig:fig3}(c)). This simulated ellipsoid Fermi pocket has its major axis along the $k_y$ ($\hat{b}$) direction due to the proximity to the HOSP zero-point of $k_y$ dispersion, resulting in a decreasing QO frequency when rotating from the $\hat{c}$ to the $\hat{b}$ axis and roughly flat values when rotating from $\hat{c}$ to the $\hat{a}$, similar to the experimental data. 

A noteworthy discrepancy of $\lesssim$0.05 kT is seen between the measurements in the right and left panels of Fig. \ref{fig:fig3}(c) for QO frequencies close to the $\hat{c}$ axis. As two different samples were used for the measurements, this can be explained within our model if the sample used for Fig. \ref{fig:fig3}(c, right) had a $\sim$0.5 meV lower chemical potential. However, it may alternatively indicate an unresolved experimental dispersion. Additionally, measurements close to the $\hat{a}$ axis are expected to include an overlapping signal from a breakdown orbit with significantly smaller effective mass \cite{EatonBreakdown}, which may be responsible for a 0.1 kT discontinuity seen at $\theta_a \sim 80^\circ$.  

\begin{figure}[ht]
    \centering
    \includegraphics[width=0.48\textwidth]{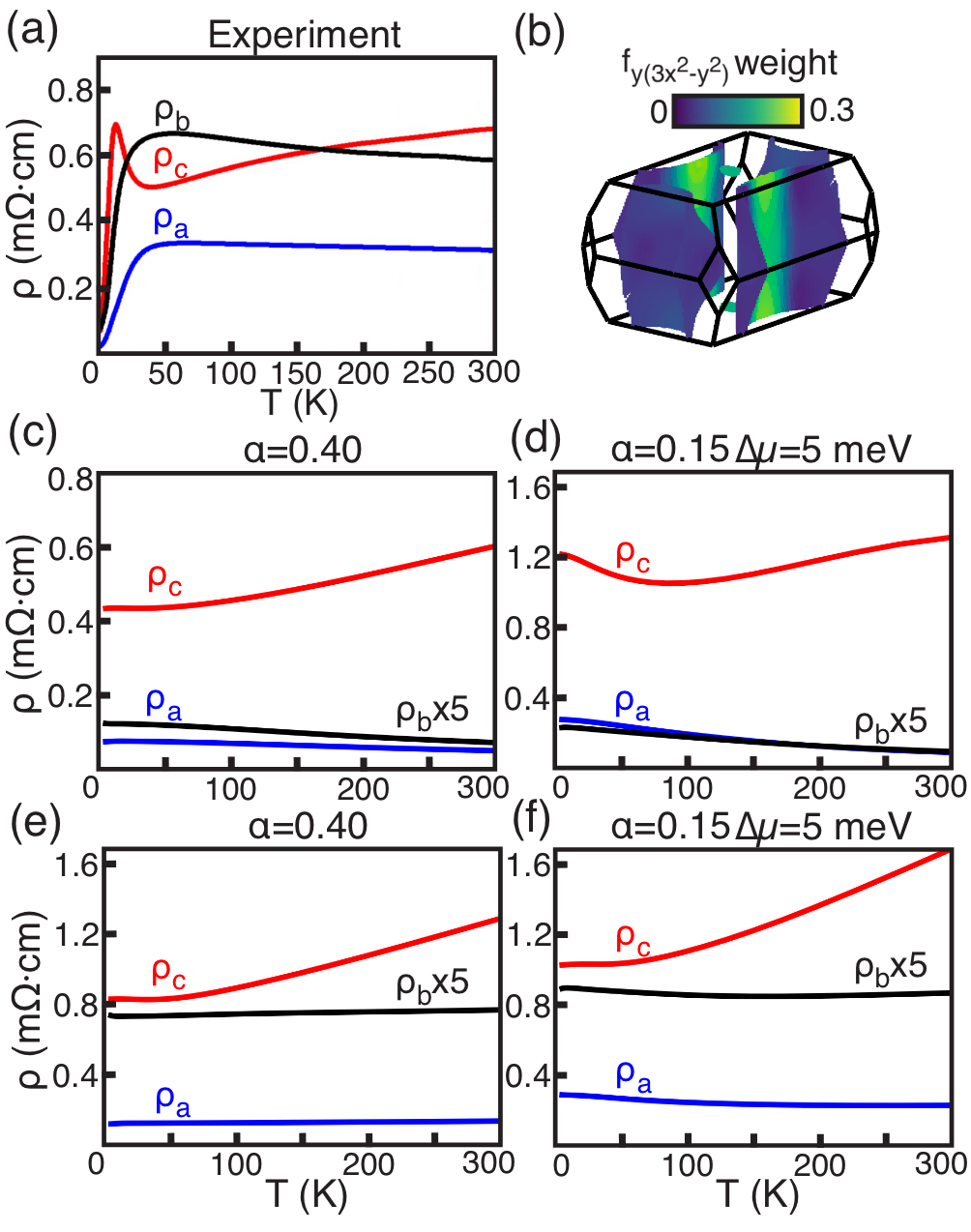} 
    \caption{\textbf{Charge transport simulation.} (a) Resistivity measurements from Ref. \cite{YunSukEo_cAxisTransport}. (b) Projected $f_{y(3x^2-y^2)}$ weight over the Fermi surface. Transport calculations from TB models with a renormalization parameter $\alpha = 0.4$ and $\alpha = 0.15$ are shown in panel (c) and (d), respectively. Panel (e) and (f) are from the same TB models, but with a momentum-dependent scattering strength.}
    \label{fig:fig4}
\end{figure}
\par
The temperature dependence of transport properties provides another basis for evaluating models of band structure near the Fermi level. Semi-classical transport calculations shown in Fig. \ref{fig:fig4}(c-f) were performed to investigate the $\rho-$T evolution driven by the thermal population of low energy band structure, assuming temperature-independent scattering strength $\Gamma(\vec{k},T)=\Gamma(\vec{k})$. While the coherence-driven transition to a Kondo metal state at $T \lesssim 50K$ is beyond the scope of this model, the monotonic higher temperature trends are a qualitative match for the experimental data, with an upward slope to $\hat{c}$ axis resistivity from 100-200K and a negative slope to resistivity within the a/b plane. This behavior arises directly from the scenario of light bands hybridizing with heavy U-5$f$ electronic states near the Fermi energy, and when 5$f$ bands are artificially removed the temperature dependence vanishes (see Fig. S5 in the SM). At low temperature, hybridization with $f$-electrons distorts the normal direction of the carrier-populated Fermi surfaces, giving electrons non-zero velocities along the $\hat{c}$ axis. At high temperature, broadening of the Fermi-Dirac distribution adds weight to less hybridized parts of the quasi-1D light bands, boosting conductivity along the $\hat{a}$ and $\hat{b}$ axes while reducing conductivity along $\hat{c}$. The $T<100K$ slope of the $\alpha=0.15$ case is an exception to this trend, as the Z-point Fermi pocket provides an additional semiconductor-type conductivity channel that counterbalances the positive correlation between $\rho_c$ and temperature Fig. \ref{fig:fig4}(d). If one includes the low temperature transition to a Kondo metal, this has the potential to create a $T<50K$ local maximum in $\rho_c$ resistivity as seen within the experimental data.

The simulations also highlight a widely known problem that the relative amplitude of resistivity along different directions does not match expectations based on the attributed band structure. Group velocities within the $k_y$-dispersive Te-5$p$ Fermi sheet are far larger than in the $k_x$-dispersive U-6$d$ Fermi sheet \cite{DMFT_LowTpocketFS,Yang_DFTandDMFT,DMFT_OrbitalSelectiveKondo,LinMiao_ARPES,Yanase_DFT1}, which results in an order-of-magnitude inconsistency with the measured resistivity ratio $\rho_b/\rho_a\sim 2$ \cite{YunSukEo_cAxisTransport} (Fig. \ref{fig:fig4}(a)). One approach to reconciling this is to posit a strong orbital symmetry dependence in the scattering rate. Figure \ref{fig:fig4}(e-f) shows simulations in which scattering strength is proportional to the $f_{y(3x^2-y^2)}$ orbital weight which has a significant impact on $\hat{b}$ axis transport due to hybridization with the Te-5$p$ Fermi sheet ($\Gamma(\vec{k})= weight(f_{y(3x^2-y^2)})\times80$ meV). This increases the $\rho_b/\rho_a$ ratio by roughly a factor of 3 but still leaves it too small by a factor of $\gtrsim 3$ and negatively impacts the temperature dependence trends. An alternative explanation might be that the highly 1D nature of the Te-5$p$ Fermi sheet leads to some degree of Anderson localization.
\par
In summary, we have started from a DFT+U simulation that reproduces the quasi-1D light electron Fermi surfaces of UTe$_2$, and added in a Gutzweiller-like renormalization term for 5$f$ electron kinetics. This is shown to convert a saddle point dispersion at the Z-point into a local minimum that provides a candidate explanation for the Z-point ARPES feature and the low frequency QO branch. The close proximity to a saddle point scenario gives this ovoid Fermi pocket a major axis along $k_y$, consistent with the angular dependence of QO. High temperature ($T>100K$) slopes of resistivity along the principal axes are qualitatively reproduced from the modeled band structure, in which they derive directly from the introduction of a 5$f$ band that disperses to within a few millielectron volts of the Fermi surface. These $T>100K$ transport trends can be reproduced on both sides of the band structure transition and do not let us distinguish between saddle point and Fermi pocket scenarios, however the occurrence of a local maximum in $\rho_c$ at $T<50K$ is potentially indicative of a 5$f$ Fermi pocket.

We stress that the paramagnetic Kohn-Sham basis of DFT does not allow a comprehensive representation of the multiplet-split electron system, and it is important not to read too much into the quantitative details of the simulation. The experimental correspondences we have identified are consistent with a basic band structure scenario in which a 5$f$ band hybridizes with the quasi-1D Te band to yield a saddle point-like band connectivity near the Fermi level, which is in turn warped into a Z-point Fermi pocket through level repulsion with a dispersive higher energy band. Within our model, the higher energy band also has predominant 5$f$ character, however the dispersion itself is derived from non-$f$ orbital symmetries for which DFT may be an effective modeling basis. (the flat underlying $f$-orbital dispersion is shown in Fig. S2 of the SM) Even if this scenario is accurate, it is entirely possible that the underlying 5$f$ band symmetries attributed by DFT+U may be incorrect.

Regardless of orbital symmetry, the presence or absence of a Z-point Fermi pocket is significant for the topology of the superconducting state \cite{AgterbergSCtopology, FuLiang_TopologicalSC} and the interpretation of emergent topological properties such as chiral edge states detected by scanning tunnelling microscopy (STM) on UTe$_2$ \cite{ChiralEdgeMode_STM}. Topological superconductivity is predicted for time reversal invariant centrosymmetric triplet superconductors that have a full superconducting gap ($A_u$ symmetry) and a Fermi surface enclosing an odd number of time reversal invariant momentum points (TRIM) \cite{FuLiang_TopologicalSC}. The symmetry of the superconducting order parameter is still under debate, however there are experiments that support an $A_u$ attribution \cite{SingleOP_Au,SingleOP_AuB1uB3u}, and it has been proposed that a Z-point $f$-electron band may result in Weyl topology in other scenarios \cite{AgterbergSCtopology}. While the well-established light band structure of UTe$_2$ encloses an even number of TRIM, our analysis shows that an additional 5$f$ derived Fermi pocket enclosing the Z-point is physically motivated in a form that can significantly improve correspondence with transport, ARPES and QO data.

\textbf{Acknowledgements:} L.A.W. acknowledges the support of the National Science Foundation under grant No. DMR-2105081.

\bibliography{ref}
\end{document}